\documentclass[12pt]{article}
\usepackage[margin=1in]{geometry}
\usepackage{setspace}
\usepackage{natbib}
\usepackage{newfloat}
\usepackage{amsmath}
\usepackage{amsfonts}
\usepackage{amssymb}
\usepackage{lineno}
\DeclareFloatingEnvironment[name={Extended Data Figure}]{suppfigure}
\usepackage[english]{babel}

\usepackage{graphicx}
\usepackage{xcolor}

\definecolor{darkgreen}{rgb}{0.0, 0.5, 0.0}
\date{}

\begin{document}

\title{Topological guidance of a self-propelled particle}

\author{
Ethan Andersson$^{1}$ and Valeri Frumkin$^{1}$\\
$^{1}$Department of Mechanical Engineering, Boston University, Boston, MA, USA
}

\begin{abstract}
Topological phenomena typically govern the behavior of delocalized waves, giving rise to robust transport in electronic, photonic, and mechanical systems. Whether similar principles can directly control the motion of a localized particle, particularly one dynamically coupled to the field that guides it, has remained largely unexplored.
Here we show that topology can govern the dynamics of a self-guided particle. Using a walking droplet whose motion is coupled to a self-generated wave field, we demonstrate that structuring the wave environment enables band-gap mediated particle exclusion, edge-guided transport, and chirality-dependent orbital dynamics arising from an emergent gauge structure. Unlike conventional topological systems, where topology constrains wave propagation alone, the present system allows global geometric structure to act directly on particle trajectories.
These results extend topological control from waves to particles and establish a route toward directing matter through global geometric design rather than local forcing.
\end{abstract}

\maketitle
Topology governs robust transport in a wide range of physical systems, from electronic materials to photonic and mechanical metamaterials~\cite{hasan_colloquium_2010,qi_topological_2011,lu_topological_2014,huber_topological_2016}. In these settings, however, topology is fundamentally a wave phenomenon: protected modes emerge from the global structure of eigenstates, while localized particles typically play no dynamical role. Whether similar principles can directly govern the motion of a particle remains largely unexplored.

Realizing such behavior requires a system in which a particle is intrinsically coupled to the field that guides it, so that global wave structure can translate into constraints on particle trajectories. The hydrodynamic pilot-wave system offers a rare setting in which a localized particle is intrinsically coupled to a delocalized wave field~\cite{bush_hydrodynamic_2020, bush_perspectives_2024}. A millimetric droplet walking on a vertically vibrated bath self-propels through resonant interaction with the surface waves it generates, forming a composite entity whose motion is inseparable from the structure of its guiding field~\cite{couder_walking_2005}. Because the droplet can propagate only where the underlying wave is supported, spatial variations in the wave environment translate directly into constraints on particle trajectories. Previous studies have shown that boundaries, obstacles, and depth variations can redirect or confine walkers, underscoring the sensitivity of their motion to the surrounding topography \cite{couder_single-particle_2006, frumkin_real_2022,frumkin_misinference_2023,primkulov_diffraction_2025}. These observations suggest that structuring the wave field at a global level may provide a route to controlling particle transport through band formation rather than local scattering.
Surface waves over shallow topography obey equations sharing the mathematical structure of electromagnetic waves in patterned media, permitting the formation of band topology within a macroscopic fluid system~\cite{chen_transformation_2009,zhu_controlling_2024,wang_topological_2025}. In the pilot-wave system, however, the consequences extend beyond wave dynamics: the droplet’s motion inherits the structure of the field to which it is coupled.

Here we show that topology can directly guide the motion of a localized particle, by exploiting the wave-particle nature of walking droplets. By engineering the droplet’s guiding wave field, we demonstrate that global topological features constrain particle trajectories, yielding band-gap mediated exclusion, edge-guided transport, and gauge-field induced orbital splitting. Together, these results establish that topology can direct the motion of individual particles through the global structure of their self-generated wave.

\subsection*{Results}
To test whether topology can constrain particle motion, we introduce structured bath topographies that shape the droplet’s guiding wave field. We find that these environments impose global constraints on particle trajectories, producing band-gap exclusion, edge-guided transport, and chirality-dependent orbital dynamics (Fig. \ref{S1}).
Experiments were conducted using walking droplets on a vertically vibrated oil bath containing submerged three-dimensional topographies. The system was operated below the Faraday threshold~\cite{faraday_xvii_1831} but above the walking threshold~\cite{couder_walking_2005}, ensuring stable wave–particle coupling. Full experimental details are provided in the Methods.

To probe band-gap mediated exclusion of a self-guided particle from a structured region, we introduce topography consisting of a rhombus-shaped bath whose interior is divided by a submerged barrier formed from a square lattice of circular pillars.    
By varying the driving frequency, we demonstrate that the lattice acts as a frequency-selective waveguide. Figure~2a shows the measured probability of droplet transmission across the submerged lattice as a function of driving frequency: at $71$ Hz the droplet crosses the submerged barrier with unit probability (supplemental video 1), while around $82$ Hz, the droplet is reflected at all times (supplemental video 2).
These observations show that band structure imposes global constraints on particle motion.
At intermediate frequencies, transmission probability drops non-monotonically with the increase in frequency. 
In the transmission regime, the droplet’s pilot wave exhibits a spatially modulated envelope whose periodicity matches that of the reciprocal lattice. Supplemental video 3 shows a close-up, slow-motion view of the modulation of the droplet's pilot-wave.
Outside the lattice, the pilot wave reverts to the familiar horseshoe-shaped profile of the unperturbed walker (supplemental video 1).  

The origin of this behavior lies in the structure of the guiding wave field. Surface waves over shallow topography obey an operator that supports band formation, so that spatial modulation of the depth imposes global constraints on wave propagation and, by extension, on particle trajectories.
In the linear regime, the surface elevation $\eta(\mathbf r,t)$ of an ideal fluid of local depth $h(\mathbf r)$, obeys 
\begin{equation}
\partial_t^2 \eta = \nabla\!\cdot\!\big[g\,h(\mathbf r)\,\nabla\eta\big]
 - \frac{\sigma}{\rho}\,\nabla\!\cdot\!\big[h(\mathbf r)\,\nabla(\nabla^2\eta)\big],
\label{eq:eta}
\end{equation}
where $g$ is gravity, $\sigma$ surface tension, and $\rho$ the density~\cite{lamb_hydrodynamics_1993}.
In the absence of capillarity, the governing equation reduces to a form equivalent to the transverse-magnetic Maxwell equation, providing a direct route to band formation.
Periodic modulation of the depth therefore imposes global constraints on the guiding field and, by extension, on the trajectories accessible to the droplet.
Capillarity introduces a higher-order correction that modifies the high-wavenumber dispersion without altering the band topology.

Having established band-gap exclusion, we next ask whether topology can guide particle motion along prescribed pathways.
To this end, we introduce a honeycomb lattice of circular pillars with broken sublattice symmetry.
We place the two domains (``lattice'' and ``anti-lattice'') side-by-side, producing a 1D interface running along the center of the bath (Fig.~3a).  
This configuration breaks mirror symmetry in opposite ways across the interface and is expected to support edge-localized modes.
When driven at frequencies outside the band gap of the bulk honeycomb crystal (here $79$ Hz), the droplet is not confined to the interface, exploring the entire lattice (Fig. 3b). However, when driven at frequencies inside the band gap of the bulk honeycomb crystal (here $83$ HZ), the droplet does not penetrate into either domain, locking onto the boundary and traveling along it (Fig. 3c). 
This guided motion demonstrates topology-induced confinement of particle transport.

Finally, we ask whether gauge structure can act directly on particle dynamics. We introduce topography consisting of an annular channel containing a chiral arrangement of submerged structures at its center.  
The spatially varying, mirror-asymmetric depth profile generates an effective gauge field for the surface waves (Fig.~4a), imparting opposite geometric phases to clockwise and counter-clockwise trajectories. 
As a result, two droplets orbiting the channel in opposite directions accumulate a measurable phase difference over time (Fig.~4b). 
This behavior mirrors the spectral splitting associated with the Aharonov–Bohm effect for charged particles encircling a magnetic flux tube, and it directly demonstrates chirality-induced gauge-field control of the walker’s orbital dynamics.

Anisotropic and chiral topographies require a tensorial description of the effective restoring force, introducing directional structure into the guiding field.
This can be achieved by generalizing the scalar depth $h(\mathbf r)$ to a symmetric second-rank tensor $\mathbf G(\mathbf r)$ which captures the directional dependence of the effective restoring force arising from layered, oblique, or spiral structures.  Here $\mathbf G(\mathbf r)$ serves as the hydrodynamic counterpart to rotated permittivity tensors in transformation optics~\cite{pendry_controlling_2006}.
With this substitution, the leading-order gravity term becomes
\[
\nabla\!\cdot\!\big[g\,\mathbf G(\mathbf r)\,\nabla\eta\big].
\]
Spatial variations of this tensor generate first-order derivative terms that can be recast in gauge-covariant form, revealing an emergent vector potential. 
Writing the operator in divergence form and isolating these contributions yields
\begin{equation}
\nabla\!\cdot\!\big[g\,\mathbf G(\mathbf r)\,\nabla\eta\big]
=
g\,(\nabla - i\mathbf A_{\mathrm{eff}})\!\cdot\!
\mathbf G(\mathbf r)\,
(\nabla - i\mathbf A_{\mathrm{eff}})\eta.
\label{eq:covariant}
\end{equation}
The explicit form of the emergent vector potential together with its detailed derivation, are given in the Supplemental Material (SM).
Equation~\eqref{eq:covariant} shows that anisotropic or chiral depth profiles act as an emergent gauge structure that acts on the droplet through its guiding field.

In a ring-like geometry, $\mathbf A_{\mathrm{eff}}$ develops a finite azimuthal component $A_\theta^{\mathrm{eff}}$, leading to discrete eigenfrequencies
\begin{equation}
\omega_m^2
= \frac{g h_0}{R^2}\bigl(m - \Phi_{\text{eff}}\bigr)^2;
\quad
\Phi_{\mathrm{eff}} = \frac{1}{2\pi}\oint_0^{2\pi} A_\theta^{\mathrm{eff}}\,d\theta,
\label{eq:AB_spectrum}
\end{equation}
where $m$ is the azimuthal mode number (see SM).  
This spectrum mirrors that of a charged particle on a flux-threaded ring due to the Aharonov–Bohm effect \cite{aharonov_significance_1959,tonomura_observation_1982}.
The degeneracy between $m$ and $-m$ is lifted whenever $\Phi_{\text{eff}}\neq 0$.
The resulting frequency splitting predicts chirality-dependent velocities, providing a direct mechanism through which gauge structure governs particle motion.

\subsection*{Discussion}
The experiments reported here demonstrate that topology can directly govern the motion of a localized particle when that particle is dynamically coupled to a guiding wave field. By structuring this field, we show that band formation can exclude a particle from a region, that topological interfaces can confine its trajectories, and that emergent gauge structure can impart chirality-dependent dynamics.

Taken together, these results establish that global wave structure can translate into deterministic constraints on particle motion. Whereas topology has traditionally been associated with extended wave states, the present system reveals how these constraints can act directly on mobile matter through wave-particle coupling.

More broadly, the intrinsic linkage between particle and field in the walker system provides a route to exploring topological transport in regimes where geometry shapes dynamics at the level of individual trajectories. Such behavior suggests new possibilities for studying topological control in systems where matter and field cannot be cleanly separated. Extending topological principles from waves to particles may enable new approaches to directing matter through global geometric design.

\section{Methods}
Our experimental setup consists of a shallow bath with 3D-printed topography, mounted on a circular base-plate with a wireless accelerometer (WitMotion WT901WIFI MPU9250 9-axis), that is driven vertically by a loudspeaker (Skaar DDX-15, 1000W).
The working fluid in all experiments was silicone oil with surface tension $\sigma=0.0209$ N/m, viscosity $20$ cSt, and density $\rho=0.965\times 10^{-3}\,\,\,\text{kg}/\text{m}^{3}$. 
For all topographies studied here, we introduced a shallow strip between the topography of interest and the confining walls of the bath, to dampen any oscillations due to the meniscus formed between the oil surface and the confining walls. The depth of the damping layer corresponded to the thickness of the oil film overlaying the highest features of the submerged topography, and was equal to $0.6\pm0.15$ mm.
We investigated three distinct classes of submerged topographies:

The first topography consisted of a rhombus-shaped bath containing a square lattice of circular pillars with radius $r=0.5$ mm, height $10$ mm, and lattice constant $a=6$ mm, used as a barrier between its two sides (Extended Data Figure \ref{S1}b). The rhombus geometry serves to steer the walking droplet away from the corners and toward the submerged barrier, ensuring consistent interaction with the lattice. The overall depth of the bath was $10.6 \pm 0.15$ mm, so that the pillars were fully submerged under an oil layer with thickness of  $0.6\pm0.15$ mm.
In that experiment, the system was driven at different frequencies for prolonged periods of time, to allow for several $10$'s of interactions between the droplet and the submerged lattice. Figure \ref{S2}
shows the ratio between the driving acceleration in the experiment and the Faraday
threshold for a given frequency. The error bars are calculated from the standard deviation of the accelerometer measurements.
For each frequency, the probability of transmission was calculated as the ratio between the number of transmission events and the total number of interactions between the droplet and the lattice. The transmission probabilities presented in Figure 2 in the main text, were computed as the ratio between the transmission events to the total number of collision events. The error bars were computed using the standard Wilson score $90\%$ confidence interval for a binomial proportion.

The second topography consisted of a honeycomb lattice of circular pillars of radius $r=0.8$ mm, with two pillar heights $a=12$ mm and $b=13$ mm, corresponding to a controlled modulation of the local depth on the $A$ and $B$ sublattices (lattice constant $c=12$ mm), producing a 1D interface running along the center of the bath (Extended Data Figure \ref{S1}c). In this experiment, the system was driven at two frequencies: $79$ Hz (outside the band-gap of the lattice), and $83$ Hz (inside the band gap), with acceleration of $87\%$ and $90\%$ of the Faraday threshold, respectively. 

The third topography consisted of a circular channel $2$ mm deep, with inner radius of $18.5$ mm and outer diameter of $24.5$ mm, enclosing a strip with chiral topography (Extended Data Figure \ref{S1}d). The well at the center is $8$ mm deep and $10$ mm in radius. For this experiment, the system was driven at $75$ Hz, with acceleration of $93\%$ of the Faraday threshold. We tracked the droplet's clockwise and counter clockwise trajectories using a standard Python particle tracking module~\cite{crocker_methods_1996}. 

In all experiments, we visualized the droplet position and pilot-wave using a charge-coupled device camera mounted above a semi-reflective mirror angled at 45° relative to the bath. We then position a diffuse-light source horizontally in front of the mirror, yielding images with bright regions corresponding to extrema or saddle points on the free surface. For the topological edge states experiment, in-place of the diffused light source, we used side illumination to increase contrast between the droplet and its surroundings, and employed standard Python particle tracking module~\cite{crocker_methods_1996} to track the droplet's trajectory.

\bibliographystyle{unsrt}
\bibliography{Topological}

\newpage

%\bibliography{sample}

\subsection*{Figures:}

\begin{figure}[h!]
    \centering
    \includegraphics[width=1 \textwidth]{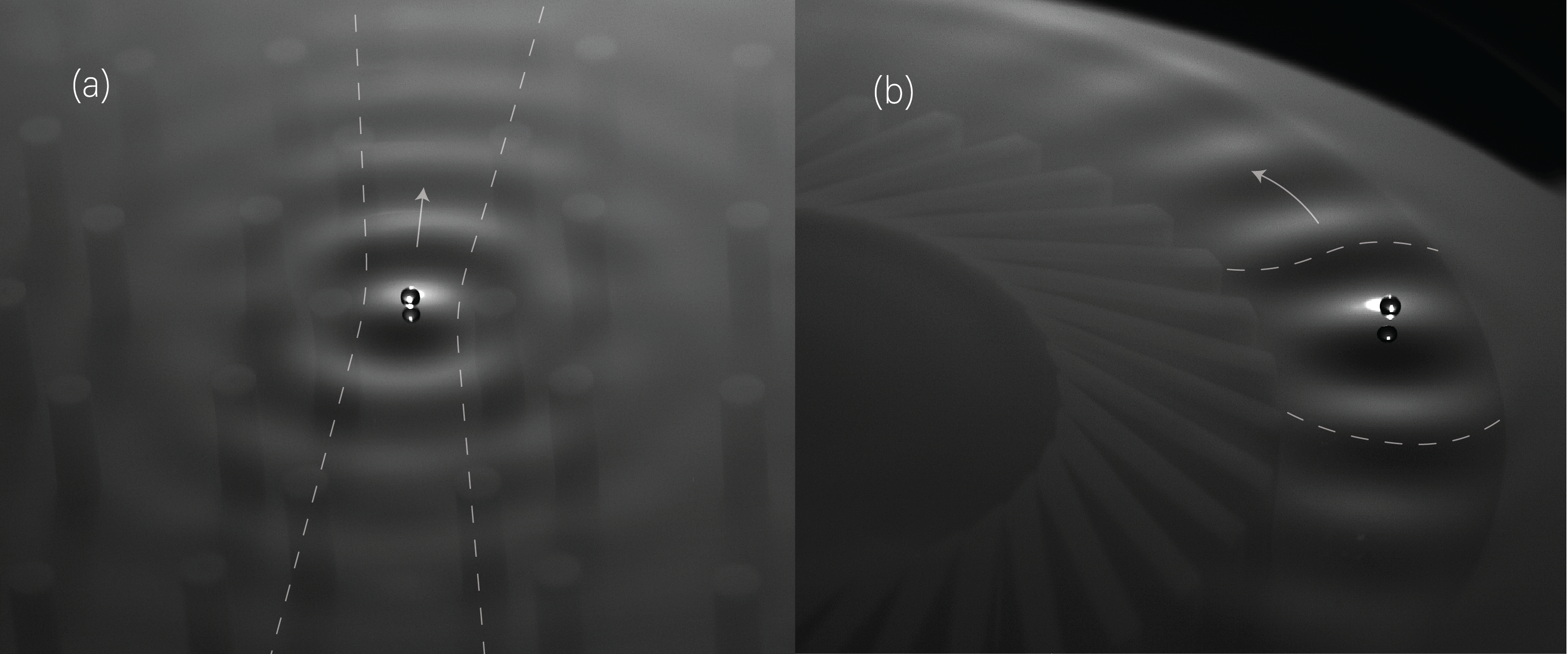}
     \caption{{\bf Topological guidance of a self-propelled particle:} (a) A droplet confined to an interface between topologically distinct domains, demonstrating edge-guided particle transport. (b)
     A walking droplet orbiting a chiral topography, exhibiting chirality-dependent dynamics arising from an emergent gauge structure.  }
    \label{Rig}
\end{figure}

\newpage

\begin{figure*}[h!]
  \centering
  \includegraphics[width=1 \textwidth]{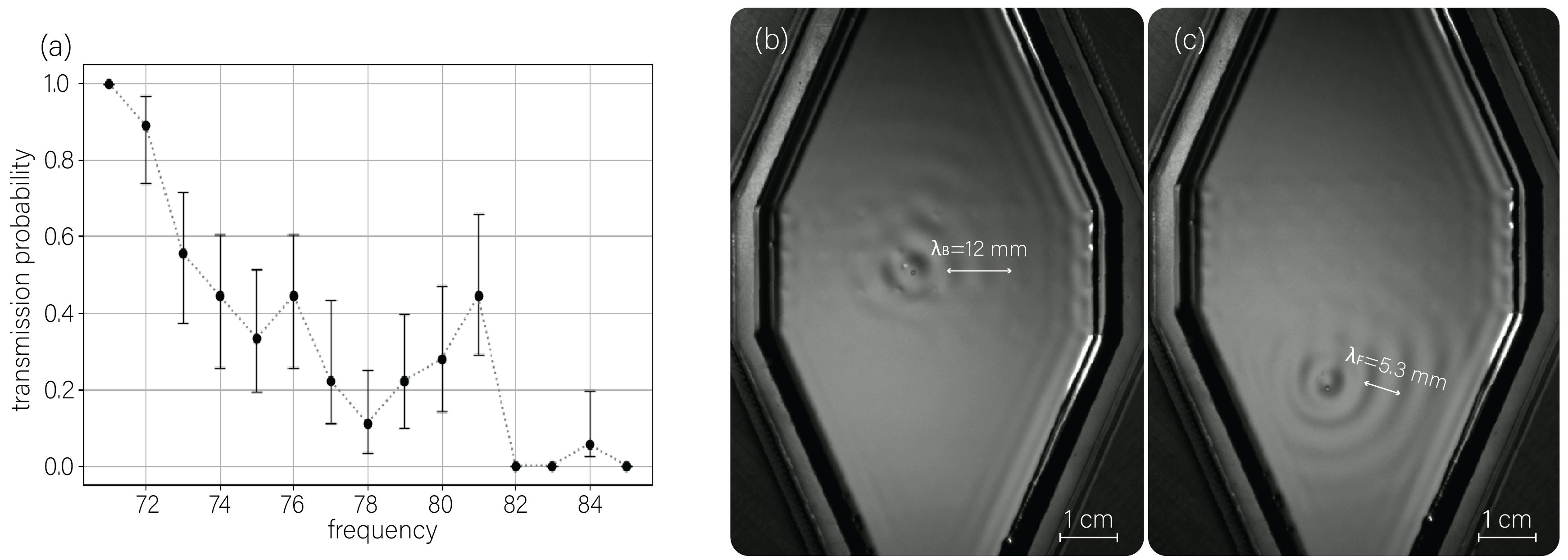}
  \caption{{\bf Band-gap mediated exclusion of a self-guided particle:} (a) Transmission probability versus driving frequency showing a transition from free propagation ($71$ Hz) to complete reflection ($82$ Hz) as the droplet encounters a band gap in its guiding wave field. (b) In the transmission regime, the particle traverses the lattice while its guiding wave adopts a spatially modulated structure imposed by the periodic topography. (c) Outside the patterned region, the guiding wave recovers its unperturbed profile, confirming that particle exclusion arises from the structured wave environment.}
  \label{fig2}
\end{figure*}

\newpage

\begin{figure*}[h!]
  \centering
  \includegraphics[width=1 \textwidth]{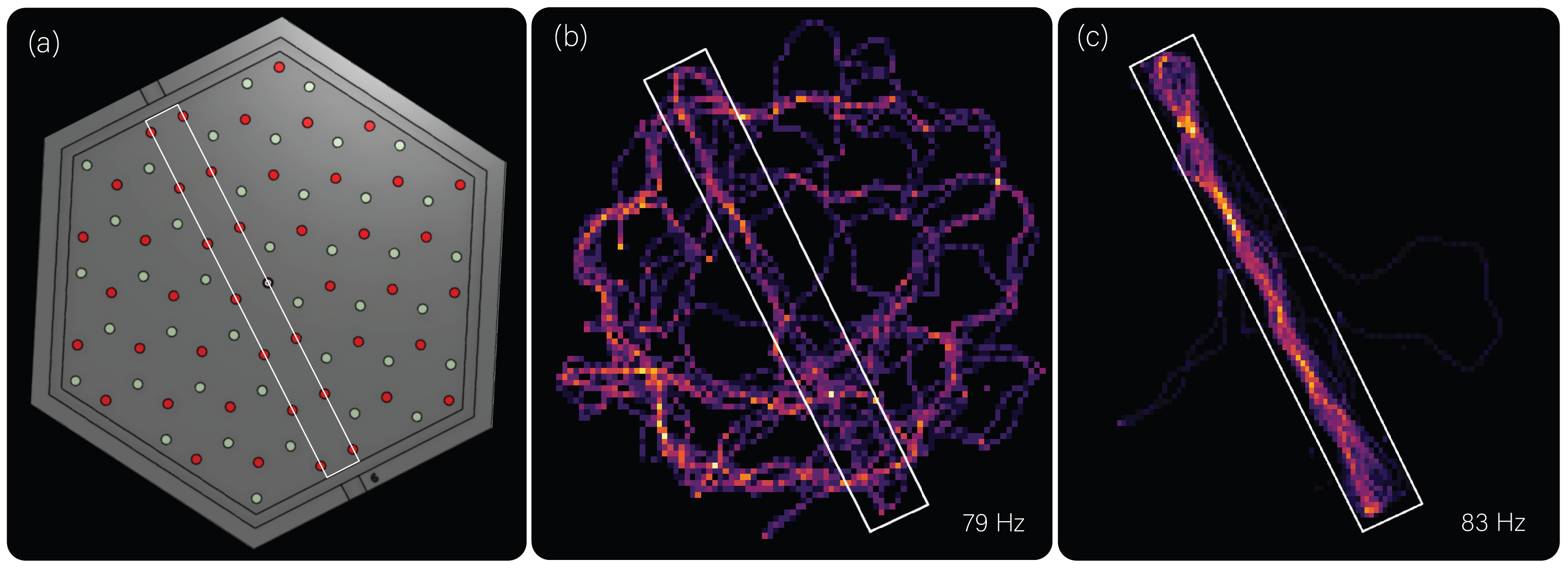}
  \caption{{\bf Edge-guided transport of a self-propelled particle:} (a) A patterned bath supporting an interface between topologically distinct domains. (b) Outside the band gap, the droplet explores the bulk lattice without directional confinement. (c) Within the band gap, the particle locks onto the interface and propagates along it, revealing topology-induced guidance of particle motion.}
  \label{fig3}
\end{figure*}

\newpage

\begin{figure*}[h!]
  \centering
  \includegraphics[width=1 \textwidth]{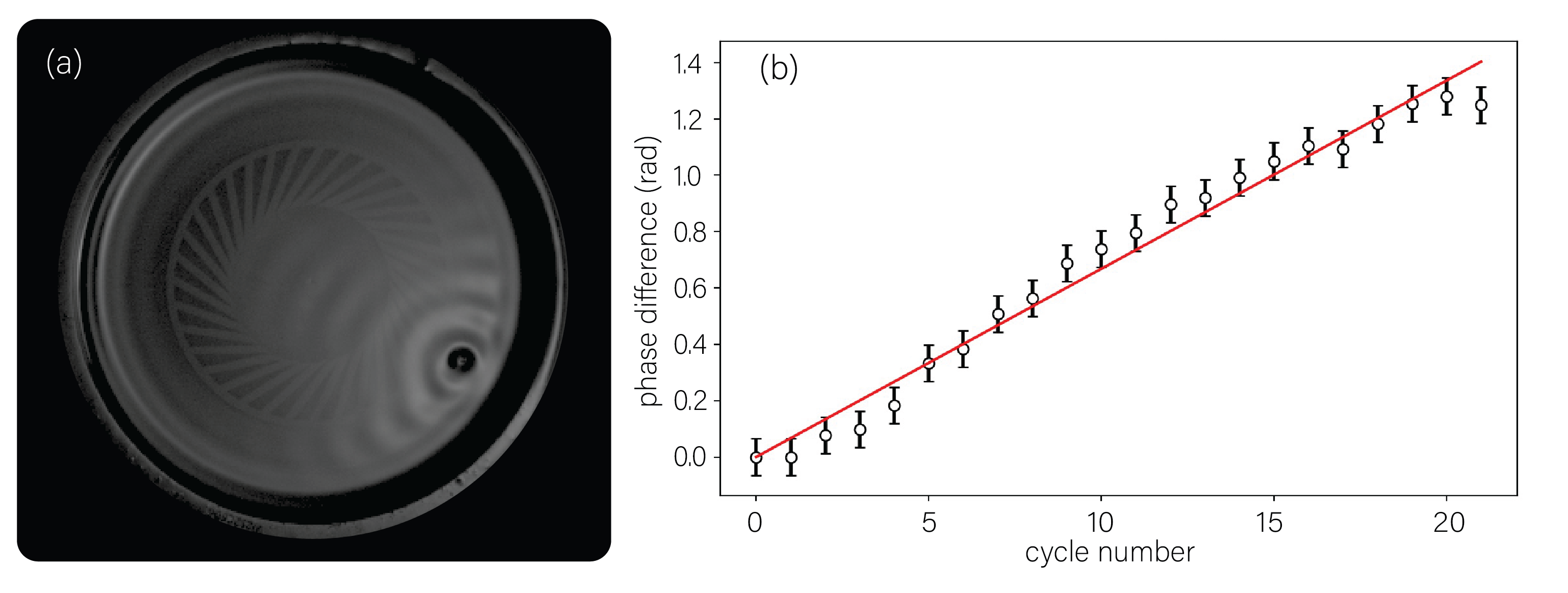}
     \caption{{\bf Gauge-field control of particle trajectories:} (a) Chiral topography skews the droplet’s guiding wave, breaking mirror symmetry in the effective dynamics. (b) Clockwise and counter-clockwise orbits accumulate different geometric phases, producing a measurable orbital splitting. The linear phase growth over multiple cycles confirms that the gauge structure acts directly on particle motion.}
    \label{fig4}
\end{figure*}

\newpage

\subsection*{Extended Data}

\begin{suppfigure}[h!]
  \centering
  \includegraphics[width=1 \textwidth]{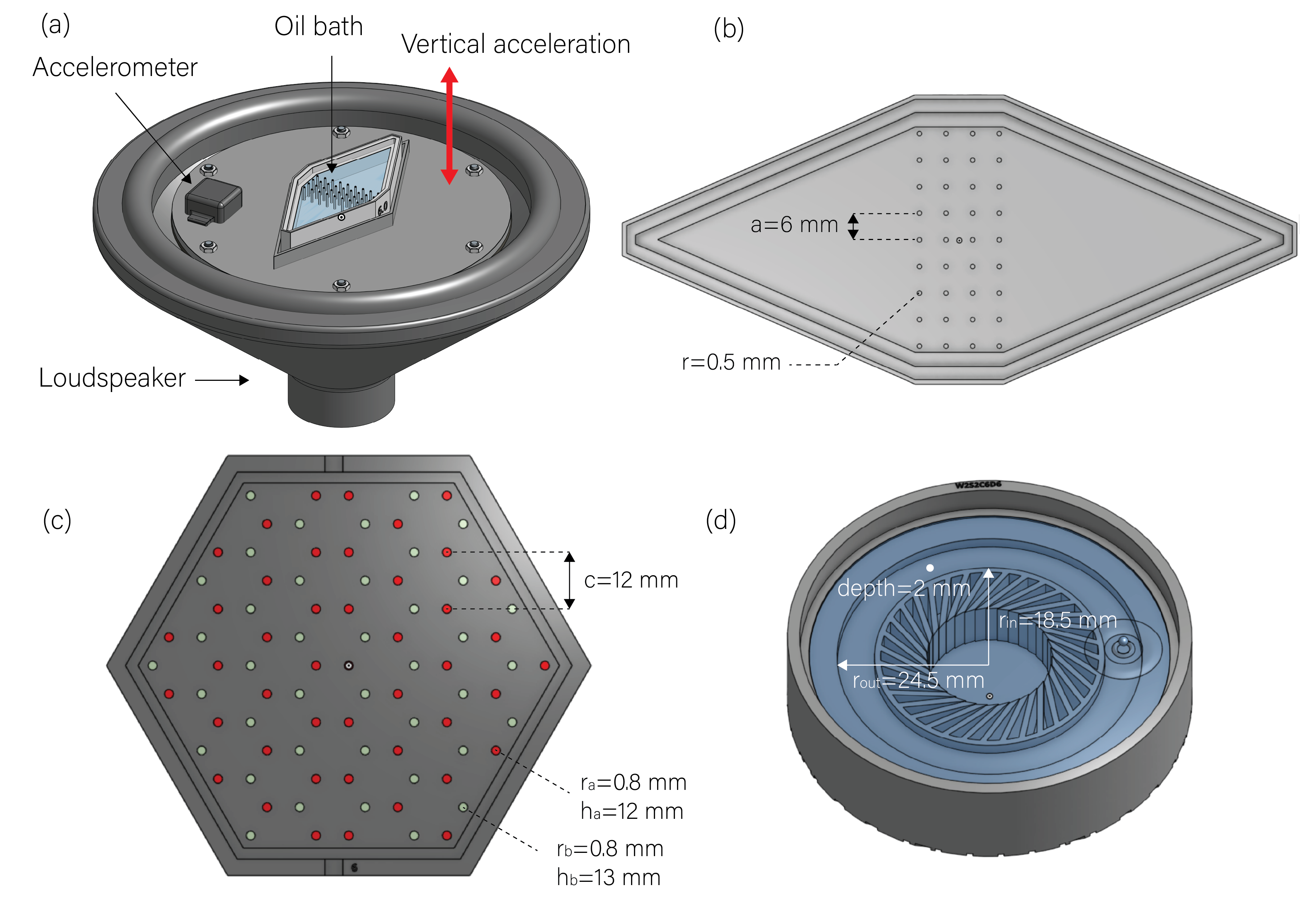}
     \caption{{\bf A schematic illustration of the experimental setup:} (a) A loudspeaker provides harmonic forcing, $f=\gamma \sin \omega t$ to a circular base plate that holds a small bath of silicon oil with 3D printed topography. (b) The topography used to study band structure: a rhombus-shaped bath containing a square lattice of radius $r=0.5$ mm and lattice constant $a=6$ mm, used as a barrier between its two sides. (c) The topography used to study topological edge states: A honeycomb lattice of circular pillars of radius $r=0.8$ mm, with two pillar heights $a=12$ mm and $b=13$ mm, corresponding to a controlled modulation of the local depth on the $A$ and $B$ sublattices (lattice constant $c=12$ mm), producing a 1D interface running along the center of the bath. (d) The topography used to study the Aharonov-Bohm splitting: A circular channel $2$ mm deep, with inner radius of $18.5$ mm and outer diameter of $24.5$ mm, enclosing a strip with chiral topography. The well at the center is $8$ mm deep and $10$ mm in radius. }
    \label{S1}
\end{suppfigure}

\newpage

\begin{suppfigure}[h!]
  \centering
  \includegraphics[width=1 \textwidth]{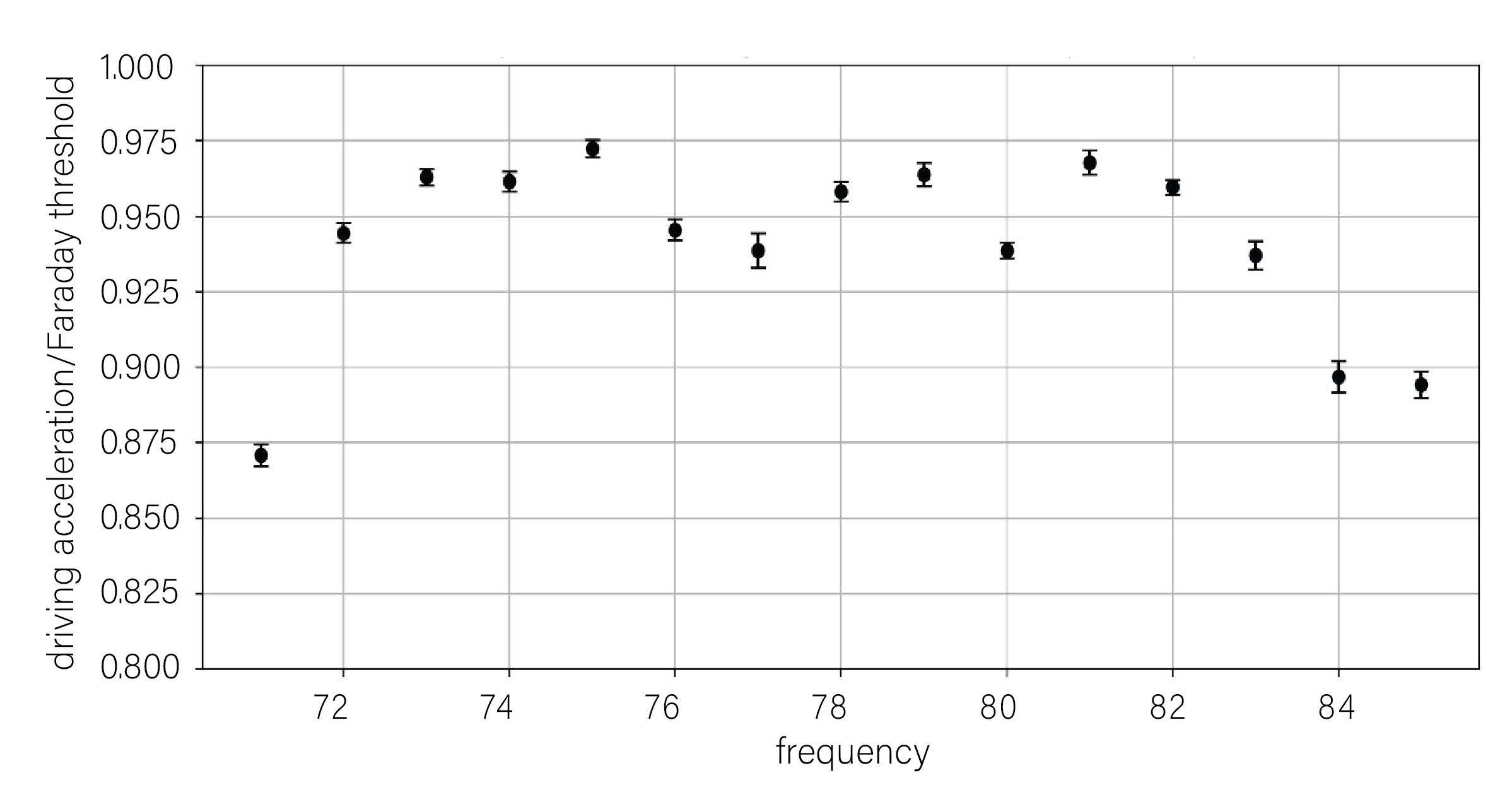}
     \caption{The ratio between the driving acceleration in the band structure experiment and the Faraday threshold for a given frequency. The error bars are calculated from the standard deviation of the accelerometer measurements.}
    \label{S2}
\end{suppfigure}

\end{document}